\documentclass[11pt]{article}

\pdfoutput=1

\usepackage{a4wide}
\usepackage{graphicx}

\begin{document}

\begin{center}   
\textbf{\LARGE Modeling the large-scale structure of a barchan dune field}

\vspace*{0.2cm}

S. \textsc{Worman}$^\natural$, A.B. \textsc{Murray}$^\natural$, R. \textsc{Littlewood}$^\flat$, B. \textsc{Andreotti}$^\sharp$ and P. \textsc{Claudin}$^\sharp$
\end{center}

{\small
\noindent
$^\natural$ Division of Earth and Ocean Sciences, Nicholas School of the Environment; Center for Nonlinear and Complex Systems, Duke University, Box 90227, Durham, NC 27708, USA.\\
$^\flat$ Unaffiliated.\\
$^\sharp$ Laboratoire de Physique et M\'ecanique des Milieux H\'et\'erog\`enes (PMMH),
UMR 7636 CNRS -- ESPCI -- Univ. Paris Diderot -- Univ. P.M. Curie, 10 rue Vauquelin, 75005 Paris, France.
}

\begin{abstract}
In nature, barchan dunes typically exist as members of larger fields that display striking, enigmatic structures that cannot be readily explained by examining the dynamics at the scale of single dunes, or by appealing to patterns in external forcing. To explore the possibility that observed structures emerge spontaneously as a collective result of many dunes interacting with each other, we built a numerical model that treats barchans as discrete entities that interact with one another according to simplified rules derived from theoretical and numerical work and from field observations: (1) Dunes exchange sand through the fluxes that leak from the downwind side of each dune and are captured on their upstream sides; (2) when dunes become sufficiently large, small dunes are born on their downwind sides (`calving'); and (3) when dunes collide directly enough, they merge. Results show that these relatively simple interactions provide potential explanations for a range of field-scale phenomena including isolated patches of dunes and heterogeneous arrangements of similarly sized dunes in denser fields. The results also suggest that (1) dune field characteristics depend on the sand flux fed into the upwind boundary, although (2) moving downwind, the system approaches a common attracting state in which the memory of the upwind conditions vanishes. This work supports the hypothesis that calving exerts a first-order control on field-scale phenomena; it prevents individual dunes from growing without bound, as single-dune analyses suggest, and allows the formation of roughly realistic, persistent dune field patterns.
\end{abstract}

\section{Introduction}
\label{intro}

Barchans are mobile, crescent-shaped dunes that form ubiquitously atop hard, flat surfaces wherever an approximately unidirectional fluid flow transports a limited amount of bed-load sediment. Barchans rarely exist in isolation but decorate desert floors, continental shelves, n\'ev\'es, and celestial bodies en masse, as members of larger fields. Despite differing granular materials and fluid mediums, barchans from these diverse environments share more than their mere characteristic crescent shape: The fields themselves also display common field-scale patterns. Fields are typically heterogeneous, featuring large areas where dunes are concentrated and others where they are absent. Where concentrated, barchans sometimes cluster into distinguishable elongated patches (Fig.~\ref{Fig1}B.1). Other times barchans congregate in expansive downwind `corridors' where dunes maintain a roughly similar size and spacing (Fig.~\ref{Fig1}B.2). However, this approximate homogeneity can be interrupted by sections of bare ground. This enigmatic structuring, with heterogeneities and homogeneities on different scales and of diverse types, is not readily explained by external forcing (Elbelrhiti et al., 2008).

In addition, despite a wealth of research explaining dynamics on the scale of single dunes, how fields of similarly sized barchans can exist at all remains an open question (e.g., Lima et al., 2002; Elbelrhiti et al., 2008; Kocurek et al., 2010; G\'enois et al., 2012). Coupled models of sediment transport and fluid dynamics explain the morphodynamic origination of individual dunes and reproduce their characteristic morphology and kinematics: Barchans emerge from the instability of a flat sand bed, march incrementally forward (via avalanching) in the direction of the mean flow at a rate inversely related to their size, and change size as a function of the relative rate that sand is obtained across their upwind-facing `backs' and lost from their downwind-facing `horns' (e.g., Kroy et al., 2002; Dur\'an et al., 2010). Given the way horn and dune width scale, larger barchans lose proportionally less sand, leading to a morphometric instability: Dunes above a certain size should tend to grow without bound, while those below should tend to shrink and disappear (e.g., Hersen et al., 2004; Dur\'an et al., 2010).

Perhaps the persistence of collections of dunes (Hersen et al., 2004; Dur\'an et al., 2010) arises as the emergent result of many dunes interacting with each other, acting as a `complex system' (e.g., Baas, 2007; G\'enois et al., 2012). Other researchers have recently modeled barchan fields as complex systems, offering the potential explanation that collections of stable dunes with similar sizes could be the result of many collisions between dunes -- events in which one dune impinges on another, culminating with two or more dunes moving away from each other (Lima et al., 2002; Diniega et al., 2010; Dur\'an et al., 2010, 2011; G\'enois et al., 2012; Hugenholtz and Barchyn, 2012). Dune-size stabilization through collisions is possible if the collisions redistribute mass in a way that tends to reduce the size of the largest dunes (Diniega et al., 2010).

Here we explore an alternative hypothesis that collections of similarly sized dunes, as well as enigmatic dune-field structures observed in nature, emerge from different dune-dune interactions: (1) the exchange of sand produced as each dune leaks sand from its horns, feeding other dunes, and (2) dune `calving' events, in which a single dune produces smaller dunes from its horns (Fig.~\ref{Fig1}A). In contrast to the collision events driving dune-field dynamics in other dune-field models (Diniega et al., 2010; Dur\'an et al., 2011; G\'enois et al., 2012), calving redistributes mass independent of dune collisions. Elbelrhiti et al. (2005) posited that calving plays a key role in stabilizing dune sizes. To explore this hypothesis, we built a numerical model that treats barchans as discrete entities that obey known empirical relationships and interact according to simplified rules derived from theoretical and numerical work and from field observations. Parameterizations herein (for numeric values not reported here, see the GSA Data Repository1) are tailored to the subaerial barchans near Tarfaya in the Western Sahara, based on extensive previous work at the dune scale (Andreotti et al., 2002) and on field-scale structuring (Elbelrhiti et al., 2008). Our goal is to present potential insights about field-scale organization, so we qualitatively describe our findings and save statistical characterizations of output for ensuing efforts.

\section{Model}
\label{model}

Barchans are typically separated from one another by bare ground so we model them as individual objects, each with a finite cross-wind width, $W$, down-wind length, $L$, and volume approximated as a pyramid, $V = W^3/40$ (Elbelrhiti et al., 2008).  In plan view, we partition $W$ into the characteristic features of barchan morphology, which are a central body defined by a slip face width, $w_s=W-w_h$, that is flanked by two horns with a combined width, $w_h$:
\begin{equation}
w_h = \Delta + \alpha W,
\label{Eq1}
\end{equation}
where $\Delta$ and $\alpha$ are numerically derived parameters (Hersen et al., 2004).

An initially empty field with a crosswind width, $Y \gg W$, and down-wind length, $X \gg W$, is resolved into square cells with sides $c = W_0/4$, where $W_0$, the width of an elemental dune, is equal to the fastest-growing wavelength that arises from the instability of a nearly flat sand bed. The algorithm below is iterated every time step, $dt$, where $dt \ll T$, where $T$ is a characteristic time related to how long it takes to transport the volume of sand equivalent to an elementary-dune volume, $V_0$: $T = W_0^2/q_{\rm sat}$, where $q_{\rm sat}$ is the `saturated flux' that a given wind regime will transport across a flat, sand-covered surface (Andreotti et al., 2002).  (For sub-aerial dunes in the Tarfaya field, $W_0 \simeq 20$~m and $T \simeq 2000$~days, see Supplementary Material).

\subsection{Sand Supplied}
At the upwind boundary, the field is supplied with a total sand flux, $q_{t,i}$, consisting of a `free flux' (wind-blown sand) $q_{f,i}$ (uniformly distributed along $Y$), and $q_{b,i}$, a `bulk flux' (migrating dunes) which represents the spontaneous formation of $n_0$ elemental dunes (Elbelrhiti, 2012), 
\begin{equation}
n_0 = \frac{q_{b,i} Y}{V_0} \, dt,
\label{Eq2}
\end{equation}
positioned randomly along $Y$.  

\subsection{Sand Transported}
All cells are then assigned a free flux, $q_f$.  Traveling considerably faster than dunes, we propagate $q_{f,i}$ instantaneously and directly downwind (neglecting lateral diffusion for simplicity) so $q_f = q_{f,i}$  until a dune is encountered, and then is modified to reflect how dune morphology impacts wind-blown transport:  Downwind of a slipface, $q_f = 0$ to mimic sand trapping due to boundary layer separation, while downwind of each horn, $q_f = q_{\rm sat}$ to capture sand leaking at the saturated rate (Hersen et al., 2004). For bulk transport, we assume no lateral migration and advance all dunes a downwind distance inversely related to $W$ through the Bagnold-like translation speed, $v$, 
\begin{equation}
v = \frac{b q_{\rm sat}}{\lambda_c + W} \, ,
\label{Eq3}
\end{equation}
where $b$ is a numeric constant and $\lambda_c$ is the length scale below which dunes disappear, included to improve accuracy (Elbelrhiti et al., 2008 and references therein). 

\subsection{Dune Volume Change}
The net sand gain/loss of a dune is calculated as the difference in sand received at a rate of $q_f$ across $W$, and lost at the saturated rate across both horns (Hersen et al., 2004): 
\begin{equation}
\frac{dV}{dt} = q_f W - q_{\rm sat} w_h.
\label{Eq4}
\end{equation}
(When dunes shrink according to (\ref{Eq4}) to the width $\lambda_c$, they are assumed to lose height until the slip face disappears; dunes below $\lambda_c$ in the model lose sand at the saturated-flux rate across their whole width.) Because barchans introduce spatiotemporal heterogeneities in $q_f$, the location of a dune relative to upwind dunes has bearing on the first term in (\ref{Eq4}). 

\subsection{Dune Collisions}
We allow for two possible outcomes when dunes collide (Hersen and Douady 2005; Katsuki et al., 2005).  If the center of mass of a dune falls within the slip face of another dune the two colliding dunes coalesce into a single composite dune that is located at their volume-weighted average position.  If two dunes touch but the above condition is not met, for example contact occurs only between their horns, they remain separate. Unlike the collision rules in other dune-field models (Diniega et al., 2010; G\'enois et al., 2012), collisions in our model do not redistribute mass in a way that tends to reduce the size of large dunes.

\subsection{Dune Calving}
\begin{figure}[t]
\centerline{\includegraphics[width=\textwidth]{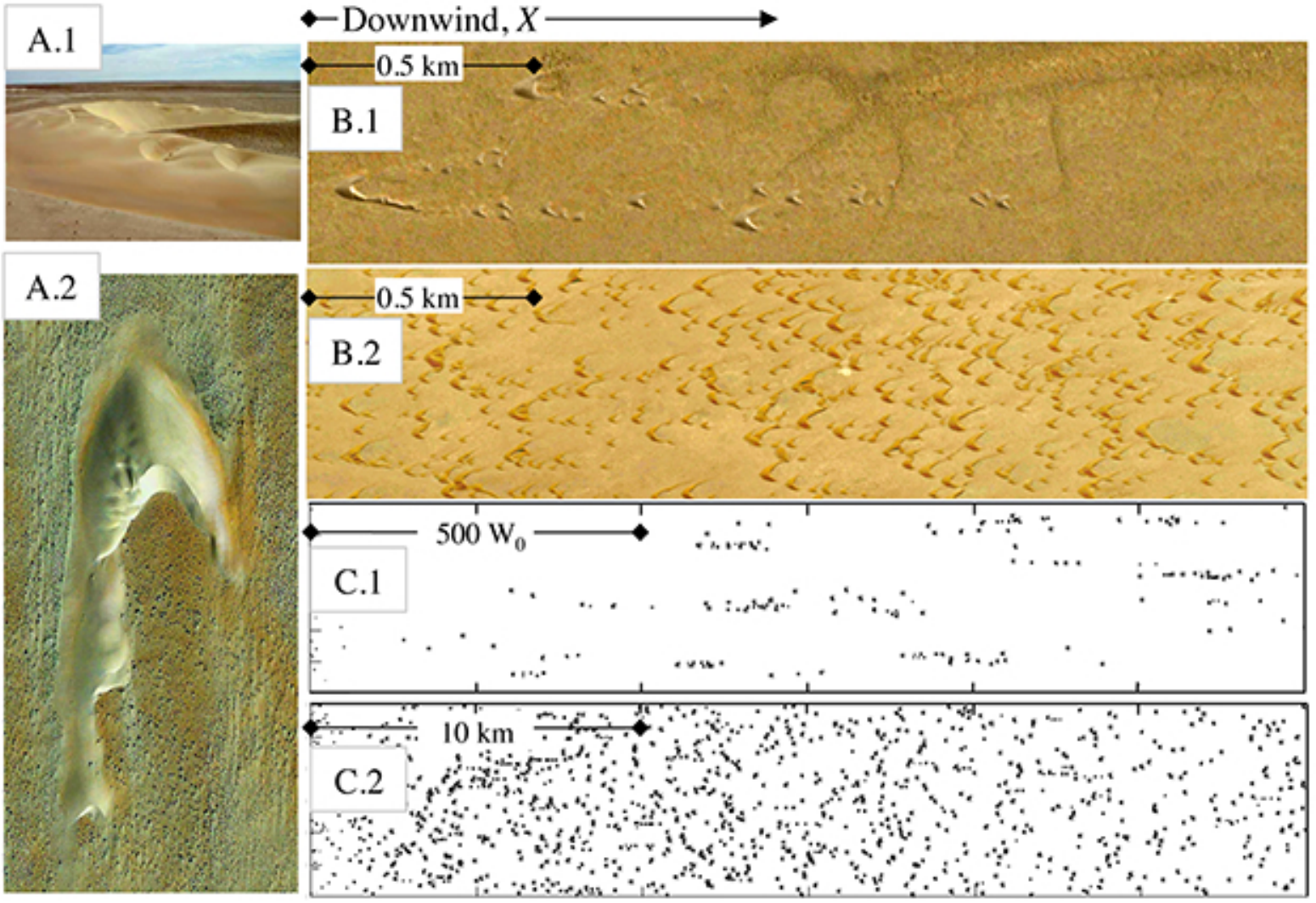}}
\caption{A: Barchan phenomena: Superimposedbedforms (A.1, courtesy of Thierry Kunicki), and a new barchan being `calved' from the horn of a `parent' dune (A.2). B: Images from Tarfaya dune field (Western Sahara) exemplifying `patches' (B.1) and `corridors' (B.2) of similarly sized dunes. C: Simulations showcasing main model results. Field patterning varies with rate that sand is supplied at upwind boundary, $q_{b,i}= 0.20 \, q_{\rm sat}$ (C1), and $q_{b,i}= 0.25 \, q_{\rm sat}$ (C2). Length scale shown in terms of elemental dune width $W_0$ on top, and in kilometers when scaled for Tarfaya field on bottom. For time series depicting approach and maintenance of steady state, see Figs.~\ref{S3}, \ref{S4}, and for a greater variety of input fluxes, see Fig.~\ref{S9}. (panels A.2 and B from Google Earth)}
\label{Fig1}
\end{figure}

The surface of a dune is subject to the same instability affecting a flat sand bed (Fig~\ref{Fig1}A; Elberhibiti et al., 2005, Zhang et al. 2010), so all modeled dunes that are sufficiently large implicitly feature superimposed bedforms.  Based on observations, surface perturbations with initial heights, $h_0$, form with a downwind spacing of $W_0$ atop both dune flanks, propagate with a speed calculated from (\ref{Eq3}), grow exponentially at a rate, $\sigma$, and are `calved' as elemental dunes if they develop a slip face (sufficient aspect ratio, $h(t)/W_0$) by the time they reach the end of their host horn (Elbelrhiti et al., 2005). All successfully calved dunes are located immediately downwind from their parent horn and are treated akin to all other dunes in subsequent time steps.

This formulation allows only dunes above a certain size, $W_c$, to produce calves. Yet dunes in higher-density areas begin calving at smaller sizes (unpublished observations), apparently because the turbulent wake of one dune exaggerates perturbations on a nearby downstream dune (Zhang et al., 2010). We account for this by letting $h_0$ assume one of two different values depending on whether a dune is located inside or outside the wake zone of an upwind neighbor (Zhang et al., 2010). We adjusted $h_0$ values so that our model dunes begin calving at widths roughly consistent with the minimum dune sizes observed to produce calves in the Tarfaya dune field: $\simeq 60$~m ($\simeq 3W_0$) and $100$~m ($\simeq 5W_0$) in `crowded' and `un-crowded' areas, respectively.

\section{Results and discussion}
\label{results}

Initial experiments revealed that output varies dramatically with total sand supply (Ewing and Kocurek, 2010; Eastwood et al., 2011) but is rather insensitive to the partitioning between the two transport modes (see Figs.~\ref{S1} and \ref{S2} in the Supplementary Material appendix).  To clearly distinguish between the consequences of external forcing and internal dynamics we hold $q_{f,i}$  at zero and vary only $q_{b,i}$, meaning we change the rate of dune seeding between simulations.  

A diverse array of patterns, similar to those observed in nature, result from the relatively simple dune interactions (Fig.~\ref{Fig1}C), consistent with the conception of barchan fields as emergent phenomena (Baas, 2007):  Simulated dunes appear in distinct spatial arrangements as a result of small-scale processes in which the large-scale effects are not obvious. In all runs, the sediment-flux conditions at the upwind boundary would cause any dune seeded there, if it remains in isolation, to vanish. Therefore, stochastic interactions between dunes are crucial to the formation of any non-empty model dune field. By releasing sand as they vanish, seeded dunes function as spatially concentrated sources of $q_f$ that can feed any dune, `luckily' located directly downwind.

\subsection{Sand Supply}
There is a threshold sand supply, $q_{b,i}$, below which simulated fields remain vacant (which varies given the probabilistic behavior of this model, as larger domains and/or longer durations both make low-frequency occurrences more likely to be observed) and beyond which fields become increasingly more populated  (Fig.~\ref{Fig1}C). 

\subsection{Low Sand-Flux and Patches}
When $q_{b,i}$ is relatively low, fields are sparse and dunes appear in spatially distinct patches that are elongated in the downwind direction (Fig.~\ref{Fig1}B and \ref{Fig1}C. As $q_{b,i}$ increases between runs and as $q_f$ increases downwind within a given run (due to disappearing dunes), patches become more prevalent and individual patches also tend to host more dunes.  This observed non-linear response to sand supply at the field-scale follows from single dune-scale dynamics as well as from how dunes operate as a collective, within a patch (see Supplementary Material).  

\subsection{High Flux Runs and Corridor-like Structuring}
\begin{figure}[t]
\centerline{\includegraphics[width=\textwidth]{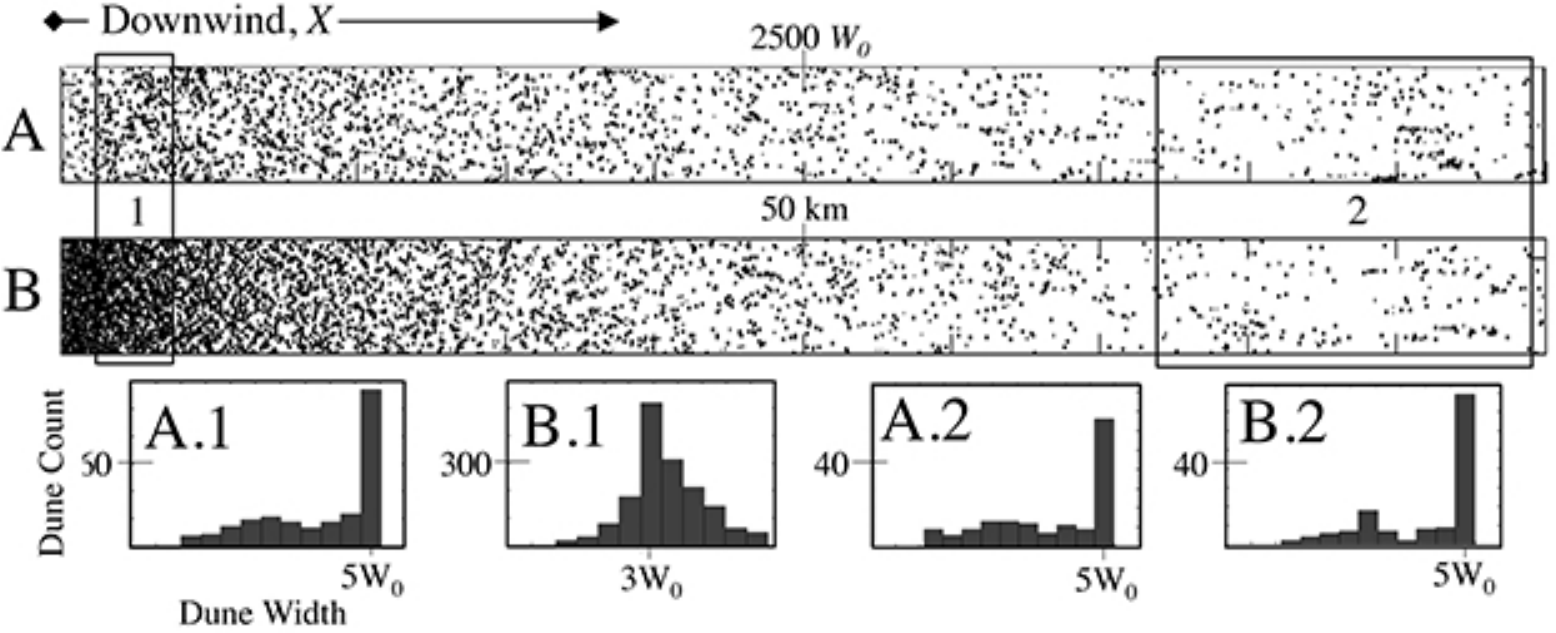}}
\caption{Dependence of field patterns on sand supply rate diminishes with distance downwind. Fields appear to approach a common attractor. A: $q_{b,i}= 0.25 \, q_{\rm sat}$. B: $q_{b,i}= 0.40 \, q_{\rm sat}$. Length scale shown as in Fig.~\ref{Fig1}. Histograms of different regions demonstrate that local dune size distributions near upwind boundary varies, with more crowded region hosting smaller dunes (see Fig.~\ref{S10} for natural field displaying this inverse relationship).}
\label{Fig2}
\end{figure}

As $q_{b,i}$ increases further, fields become more homogeneous in the sense that dunes are observed throughout the domain (Fig.~\ref{Fig1}C), and locally within a given region, display a characteristic size (Fig.~\ref{Fig2}). Bare areas are interspersed among the otherwise continuous sprinkling of dunes so these fields also possesses some degree of spatial heterogeneity (Fig.~\ref{Fig1}C). Given this display of features also common to natural barchan corridors, we refer to the patterns resulting from these higher-flux runs as simulated corridors.

\subsection{Upwind Dependence on Input Flux, Downwind Independence}
After dunes migrate to the downwind end of the domain, all simulated fields attain a statistically steady state (with the exception of the occasional development of mega-dunes, see Figs.~\ref{S5} and \ref{S6} in the Supplementary Material appendix).  Despite persisting differences in the upwind portion of the domain for runs with different $q_{b,i}$, downwind portions of fields are strikingly similar (Fig.~\ref{Fig2}):  The dune-dune interactions, when spatially iterated between successive dunes and collections of dunes over a sufficient distance, appear to lead toward a common attractor.  Although most natural fields lack these expansive downwind lengths, convergent behavior could nonetheless be tested for with empirical work focused on down-wind field trends which, unlike transverse structure (Elberhiti et al., 2008), have not yet been systematically examined.

\subsection{Dune Calving Within a Field}
This work supports the hypothesis that calving plays a fundamental role in field-scale phenomena (Elbelrhiti et al., 2005). Because the collision rules in our model do notfavor the stabilization of dune sizes, collisions play no direct role in overcoming the morphometric instability as they do in other models (Diniega et al., 2010; G\'enois et al., 2012) -- in fact, collisions in our model can produce an anticipated instability leading to runaway growth as an anomalously large dune moves more slowly than upstream neighbors (e.g. Hersen et al., 2004), which occurs very rarely given the regional consistency of dune sizes and therefore migration rates (Figs. \ref{S5} and \ref{S6}). Although we believe our collision rules are physically reasonable, we have chosen not to construct rules that would contribute to stabilizing dune sizes, to increase the clarity of our numerical experiments testing the importance of inter-dune sand exchanges through free flux and calving. When the calving process is disabled, dunes grow without bound and fields lack any resemblance to natural ones (Figs. \ref{S7} and \ref{S8}).  When enabled, calving -- a sand loss term unaccounted for in Eq.~\ref{Eq4}  -- prevents this runaway growth and allows patterns to develop which maintain a roughly realistic appearance. In our model, based on field observations (Elbelrhiti et al., 2005) and numerical modeling (Zhang et al., 2010), imminent dune collisions can produce calving, as the perturbation to the wind field caused by an upstream dune enhances the formation of superimposed bedforms on a nearby dune. Thus, this work suggests that collisions help stabilize dune sizes through their effect on calving.

In our model, although a dune much larger than $W_c$ can calf as frequently as successive waves can migrate to the end of their host horn, as a growing dune reaches $W_c$ and ejects a calf, its volume is diminished and its width drops below $W_c$, interrupting the process. After recovering enough sand to again surmount $W_c$, it then calves and ceases calving once again. In the model, this negative feedback keeps dunes within a narrow width range and leads to what, locally at the community scale, is a `single' stable size (Fig.~\ref{Fig2}). Moreover, the time between two successive calving events depends on $W$: Larger dunes recover sand more quickly and therefore also release it more rapidly downwind by calving at a higher rate. Additionally, because the sand supply within a field is inherently stochastic, calving is also stochastic and is moderated by the behavior of neighboring dunes. For example, noting that dunes are typically centered on the horn of an upwind neighbor, a dune can absorb a calf too -- which would cause a dune just below $W_c$ to become large enough to calve again. Therefore Equation~\ref{Eq4} is only loosely applicable to dunes in structured fields where the rate of sand gains and losses appears coupled. Short-term empirical measurements grounded our formulation of the calving process (Elbelrhiti et al., 2005), and longer-term empirical observations, for example on the relationship between calving frequency and dune size and field density, is needed to evaluate whether or not this calving-frequency-regulation mechanism occurs in natural dune fields too.

The correspondence between the characteristic dune size of a region and $W_c$ -- either the crowded or un-crowded value (Figs.~\ref{Fig2} and \ref{S10}) -- also points to calving as the model mechanism regulating dune size.  Moreover, the emergent, inverse relationship between the size and density is consistent with the structuring observed in the Tarfaya dune field (Fig.~\ref{Fig2}; Elbelrhiti et al., 2008).  If this mechanism regulates dune size in nature, it is still unclear how different corridors hosting different sized dunes, and therefore by inference different calving sizes, can be juxtaposed within the same field. The only plausible explanation offered by this model is regional variation in sand supply. This suggestion is consistent with field data (Elbelrhiti et al., 2008) if they are reinterpreted in light of the non-linear response of the model system to variations in input flux:  The differences in regional sand fluxes within the Tarfaya field ($\simeq 0.05 q_{\rm sat}$) are significant enough to cause different characteristic dune sizes in the model, near the sand source (i.e. in the upwind portions of the modeled fields; Fig.~\ref{Fig2}) and could result from regional variations in lithology, topography, or fluxes at the field boundaries.  

\subsection{Sensitivity to Parameters and Robustness of Potential Insights}
Although some of the parameterizations in this exploratory modeling effort are not tightly constrained, the results presented here arise from what we believe are the most reasonable choices among the simple formulations we have experimented with. In addition, as long as calving is incorporated in the model, varying most parameters alters the range of input fluxes that produce the various patterns and behaviors, but does not fundamentally change the qualitative results we have emphasized here. In particular, the emergence from internal dynamics of spatial heterogeneity within a field -- including patchiness at relatively low input fluxes and variations in dune density with higher sediment supplies -- results robustly from combinations of parameterizations we have tried.

\section{Conclusions}
\label{conclu}

The numerical modeling results show that a rich array of dune field-scale phenomena could be emergent results of collections of barchan dunes interacting with each other in relatively simple ways. Dune-field patterns, including spatially intermittent patterns with characteristic elongated structures, and denser corridors of similarly sized dunes with internal spatial heterogeneity, arise from two main dune-interaction mechanisms: (1) the exchange of sand via free flux (lost from the horns and intercepted on the upwind flanks); and (2) the calving of small dunes from the horns of sufficiently large parent dunes. The dune collision rules in this work do not play a role in counteracting the morphometric instability predicted on the basis of single-dune models, showing that the existence of collections of stable-sized dunes could result from calving, rather than from collisions. Changing the flux of sand input at the upwind boundary of the domain affects the field-scale patterns, dune density, and characteristic sizes of the dunes -- although the dune interactions, when iterated far enough downwind, cause the system to approach a common attractor (patchy, relatively sparse dunes), independent of the conditions at the upwind end of the domain.

\vspace*{0.3cm}
\noindent
\rule[0.1cm]{3cm}{1pt}

We thank ANR Zephyr grant $\#$ERCS07\underline{\ }18 for funding.


\newpage
\appendix

\section{Supplementary Material}
\label{SM}

\subsection{Model}
Numeric values of parameterizations used to simulate Western Sahara barchans, listed in order of appearance in text and drawn from Hersen et al. (2004), Andreotti et al. (2002), Elbelrhiti et al. (2005), and field observations of B. Andreotti and P. Claudin (denoted by $^*$)

\begin{center}
\begin{tabular}{|c|c|}
\hline
Parameter			& Numeric value 				\\ \hline
$\alpha$				& $0.05$						\\ \hline
$\Delta$				& $4.65$ m					\\ \hline
$W_0$				& $20$ m						\\ \hline
$q_{\rm sat}$			& $0.2$ m$^2$/day				\\ \hline
$W_\infty$			& $\simeq 30$ m $\simeq 1.5 W_0$	\\
(from (\ref{Eq4}), assuming $q_f = 0.2 q_{\rm sat}$)	&		\\ \hline
$T=W_0^2/q_{\rm sat}$	& $2000$ days					\\ \hline
$dt$, where $dt \ll T$	& $1$ day						\\ \hline
$b$					& $45$						\\ \hline
$\lambda_c$			& $16.6$ m $\simeq 0.75 W_0$	\\ \hline
$\sigma$				& $0.015$ m/day				\\ \hline
$h_0$, outside wake zone$^*$	& $0.05$ m, so $W_c \simeq 100$ m $\simeq 5 W_0$  \\ \hline
$h_0$, inside wake zone$^*$		& $0.165$ m, so $W_c \simeq 60$ m $\simeq 3 W_0$  \\ \hline
$h(t)/W_0$ (aspect ratio)			& $0.1$  \\
\hline
\end{tabular}
\end{center}
%

\subsection{Results}
\textit{Dune Interactions Within a Field}\\
Setting $dV/dt = 0$ in (\ref{Eq4}) reveals one unstable equilibrium dune width, $W_\infty$,  
\begin{equation}
W_\infty = \frac{q_{\rm sat} \Delta}{q_f - \alpha q_{\rm sat}} \, ,
\label{Eq5}
\end{equation}
that depends on $q_f$ (Hersen et al., 2004).  The spatial variation in $q_f$ caused by upstream dunes (either persistent dunes or small ones that disappear and become free flux) leads to more than one $W_\infty$, potentially permitting divergent behavior of equivalently sized dunes; `lucky' ones, with high $q_f$ from upwind, can grow, while others shrink and disappear.

\vspace*{0.2cm}

\noindent
\textit{Calving Dynamics}\\
Analytically, recently calved dunes evolve volumetrically according to,
\begin{equation}
\frac{dV}{dt} = 0.5 q_{\rm sat} [\Delta + \alpha W_c] - q_{\rm sat} [\Delta + \alpha W_0],
\label{EqS1}
\end{equation}
where the first term in (\ref{Eq4}) has been modified to represent sand supplied to a calf by one horn of its parent.  Setting $dV/dt = 0$ and solving for the parent width that produces stable calves ($W_c \simeq 130$~m~$\simeq 6.5 W_0$) implies that parent dunes, as parameterized, cannot support a calf through sand-flux exchange alone.  

The first term in (\ref{EqS1}) represents the amount of sand leaked to a calf from a parent.  After that parent has a subsequent calf, the amount of sand leaked downwind to its earlier calf, is simply the rate at which the calf loses sand, or $q_{\rm sat} [\Delta + \alpha W_0]$, which is less than the sand influx term in (\ref{EqS1}) and leads to the conclusion that subsequent calves detrimentally disrupt the sand supply of an earlier calf.

\vspace*{0.2cm}

\noindent
\textit{Patch Formation, and Interactions Within Patches}\\
The single dune located at the upwind apex of a patch initiated its formation and subsequently regulates its expansion.  Patches begin when a dune stochastically captures enough sand from disappearing upwind dunes to surmount $W_\infty$.  This pioneer then grows to the calving width, $W_c$, and ejects calves.  However, the sand balance of these progeny is negative (see \emph{Calving Dynamics} above), implying that calf survival is tied to stochastic fluctuations in sand flux in a manner similar to seeded dunes.   

Yet unlike a seeded dune, a calf is mostly sheltered from variations in $q_f$ by its parent.  Instead of buffering against fluctuations, however, a parent transmits them downwind by calving intermittently.  Once a parent calves again, this subsequent calf disrupts the sand supply of the extant calf, increasing its net rate of sand loss. While an upwind calf accentuates the sand-deficit of a down-wind calf, their relative spacing and sizes sometimes allows for their coalescence. This enables some calves to grow downwind of a pioneer horn, slowing their migration and allowing them to capture additional calves.  If they become sufficiently sized, they too begin calving and some of their progeny, likewise, go on to form another generation centered on their horns.  In this manner, model patches expand laterally and elongate in a stochastic fashion as they travel downwind.

The rate of patch initiation and expansion is controlled by $q_{b,i}$.  As $q_{b,i}$  increases, the probability that seeded dunes are successfully established increases.  Conceptually, this is a conditional probability where the rate of dune injection influences the probability, $P$, that an isolated dune gains an upwind neighbor,  
\begin{equation}
P = n_0 \, \frac{W}{Y} \, .
\label{Eq6}
\end{equation}
The additional dependence of $P$ on $W$ reveals a positive feedback:  If a dune acquires an upwind neighbor, it grows and increases its chances of acquiring another upwind neighbor and growing even further.  As a consequence, as $q_{b,i}$ increases more dunes successfully surmount $W_\infty$ and proceed to found more rapidly expanding patches.

\vspace*{0.2cm}

\noindent
\textit{Megadune Formation}\\
Although dunes producing calves tend to be approximately the same size, and therefore collide infrequently, absorbing smaller dunes can occasionally lead to a positive feedback: An unusually large dune will absorb dunes more and more rapidly as it grows (Figs.~\ref{S5} and \ref{S6}). Such very infrequent events interrupt the otherwise homogeneous pattern in the vicinity of the growing `megadune', by starving downwind dunes of sand and creating a vacant, downwind area (Figs.~\ref{S5} and \ref{S6}). In nature, unusually large `mega dunes' can occur within a dune field, although  the Atmospheric Boundary Layer (Andreotti et al., 2009) ultimately constrains the size of these dunes. Megadune formation in our model, although rare, is likely exaggerated by our conservative formulation of the calving process.  In nature, mega-dunes are sufficiently wide that multiple wave trains flank each horn (Elbelrhiti et al., 2008), an effect not included in our model parameterizations.  This presumably increases their calving rate and if accounted for would tend to retard such runaway growth (as would interactions with the atmospheric boundary layer, mentioned in the main text). 

Unusually large barchans in nature are sometimes assumed to form at the upwind boundary of a dune field, and then are used to back calculate the relative age of the dune field, dividing their distance from the upwind boundary by their propagation velocity.  This estimation then provides a rough proxy for the length of time that external conditions have been constant (Hesse, 2009).  However, in our model these large dunes form rapidly in place, far downwind of the upwind boundary, suggesting that large-dune migration rates and positions could be misleading indicators of field age.  If field-age estimates are systematically too large, this result may have implications for paleo-environmental interpretation, especially for extra-terrestrial settings where other proxies are unavailable.

\vspace*{0.3cm}

\noindent
\textbf{Additional References} (cited here in Supplementary Material but not in main text)\\
Andreotti B, Fourri\`ere A, Ould-Kaddour F, Murray AB, Claudin P (2009)\\
\textit{Giant Aeolian dune size determined by the average depth of the atmospheric boundary layer.}\\
Nature \textbf{457}, 1120--1123.

\vspace*{0.1cm}

\noindent
Hesse R (2009)\\
\textit{Do swarms of migrating barchans dunes record paleoenvironmental changes? -- A case study spanning the middle to late Holocene in the Pampa de Jaguay, southern Peru.}\\
Geomorphology \textbf{104}, 185--190.

\begin{figure}[p]
\centerline{\includegraphics[width=\textwidth]{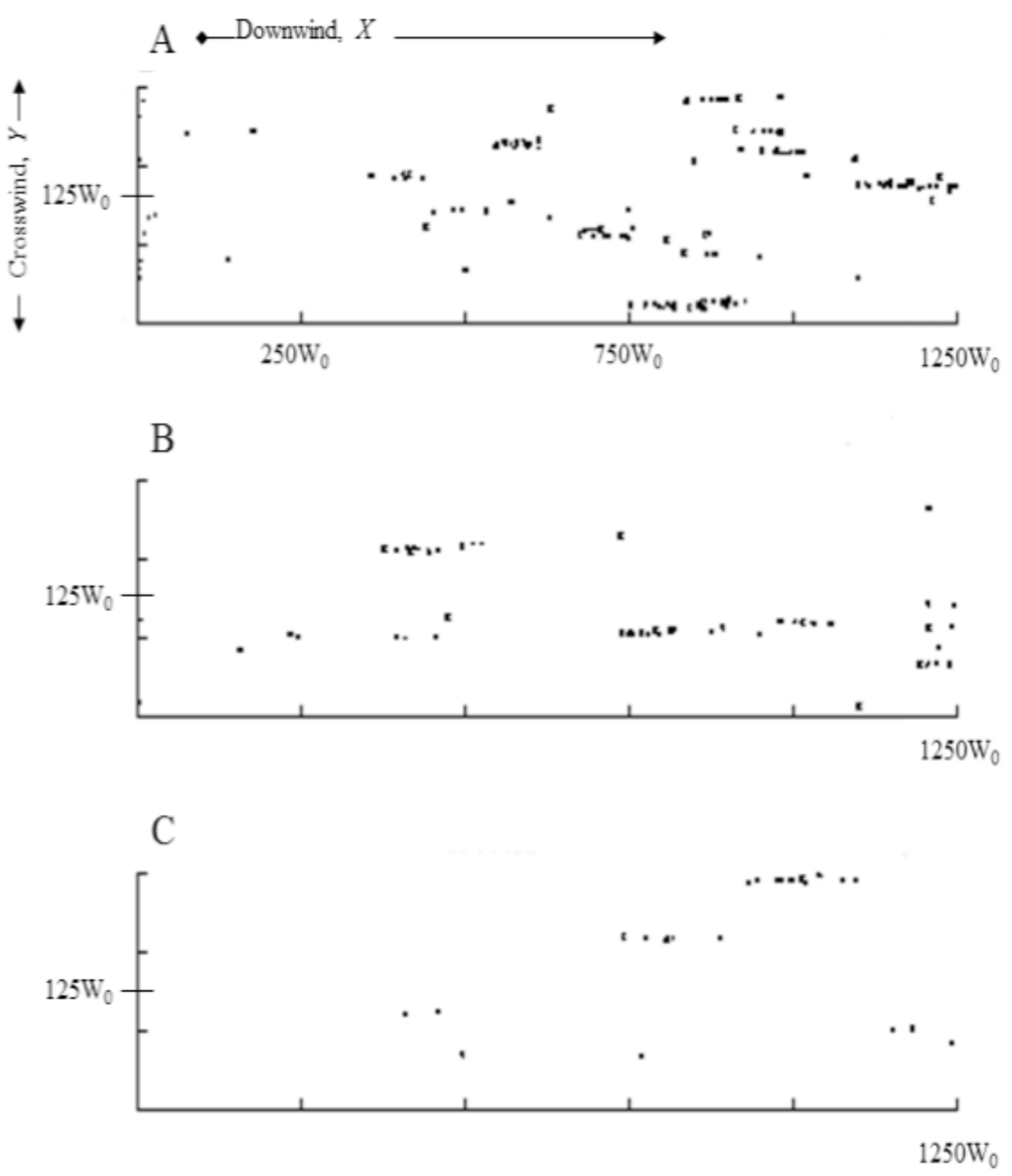}}
\caption{Demonstration of the insensitivity of model output to flux partitioning input for lower flux runs where $q_{t,i} = 0.20 q_{\rm sat}$.  Note that the qualitative patchy pattern is unchanged where (A) $q_{f,i}  = 0.10 q_{\rm sat}$ and $q_{b,i}  = 0.10 q_{\rm sat}$ (B), $q_{f,i}  = 0.05 q_{\rm sat}$ and $q_{b,i}  = 0.15 q_{\rm sat}$ and (C) $q_{f,i}  = 0.0 q_{\rm sat}$ and $q_{b,i}  = 0.20 q_{\rm sat}$.}
\label{S1}
\end{figure}
\begin{figure}[p]
\centerline{\includegraphics[width=\textwidth]{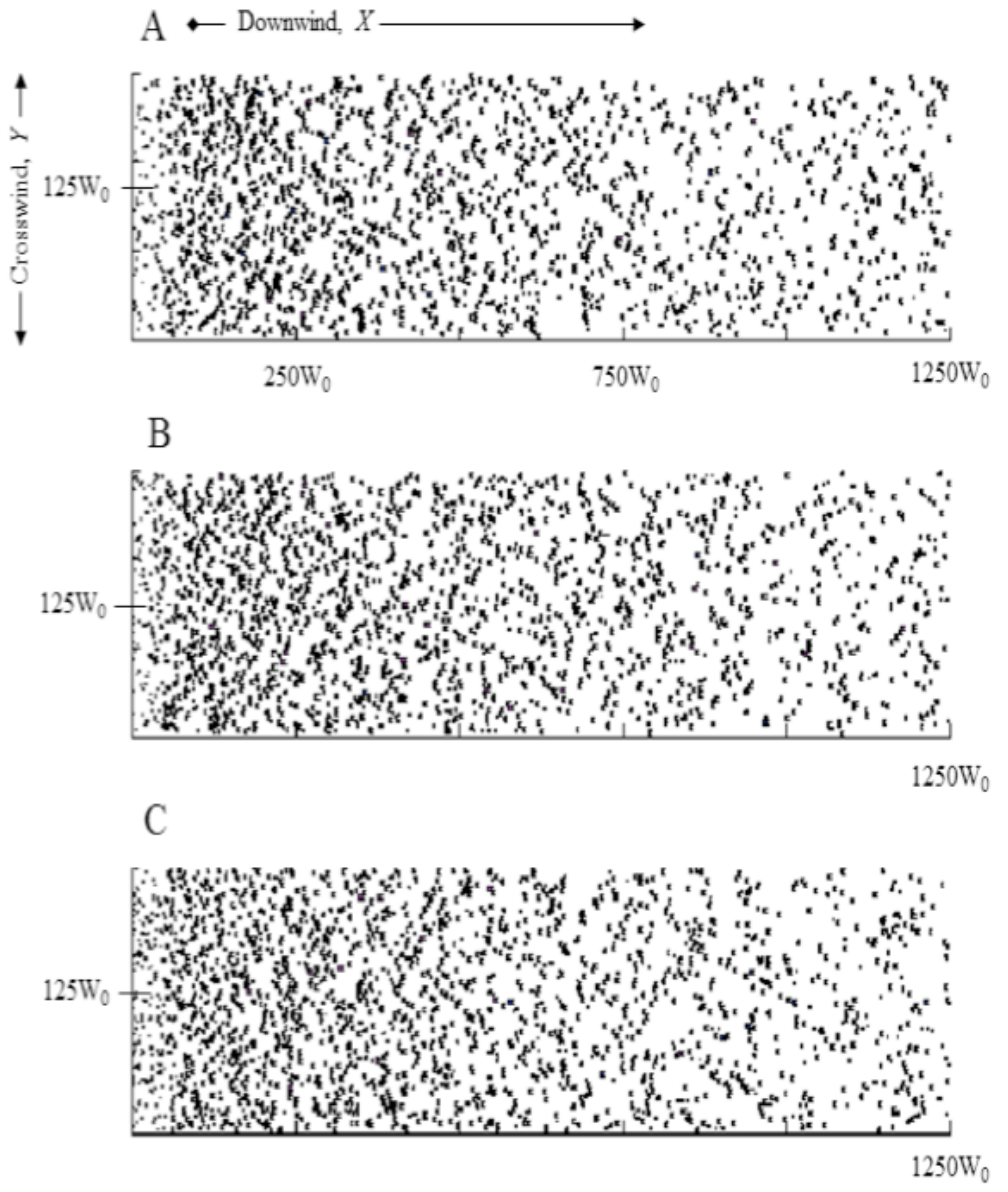}}
\caption{Demonstration of the insensitivity of model output to flux partitioning input for higher flux runs where $q_{t,i} = 0.30 q_{\rm sat}$.  Note that the qualitative corridor pattern is unchanged where (A) $q_{f,i}  = 0.10 q_{\rm sat}$ and $q_{b,i}  = 0.20 q_{\rm sat}$ (B), $q_{f,i}  = 0.05 q_{\rm sat}$ and $q_{b,i}  = 0.25 q_{\rm sat}$ and (C) $q_{f,i}  = 0.0 q_{\rm sat}$ and $q_{b,i}  = 0.30 q_{\rm sat}$.}
\label{S2}
\end{figure}
\begin{figure}[p]
\centerline{\includegraphics[width=\textwidth]{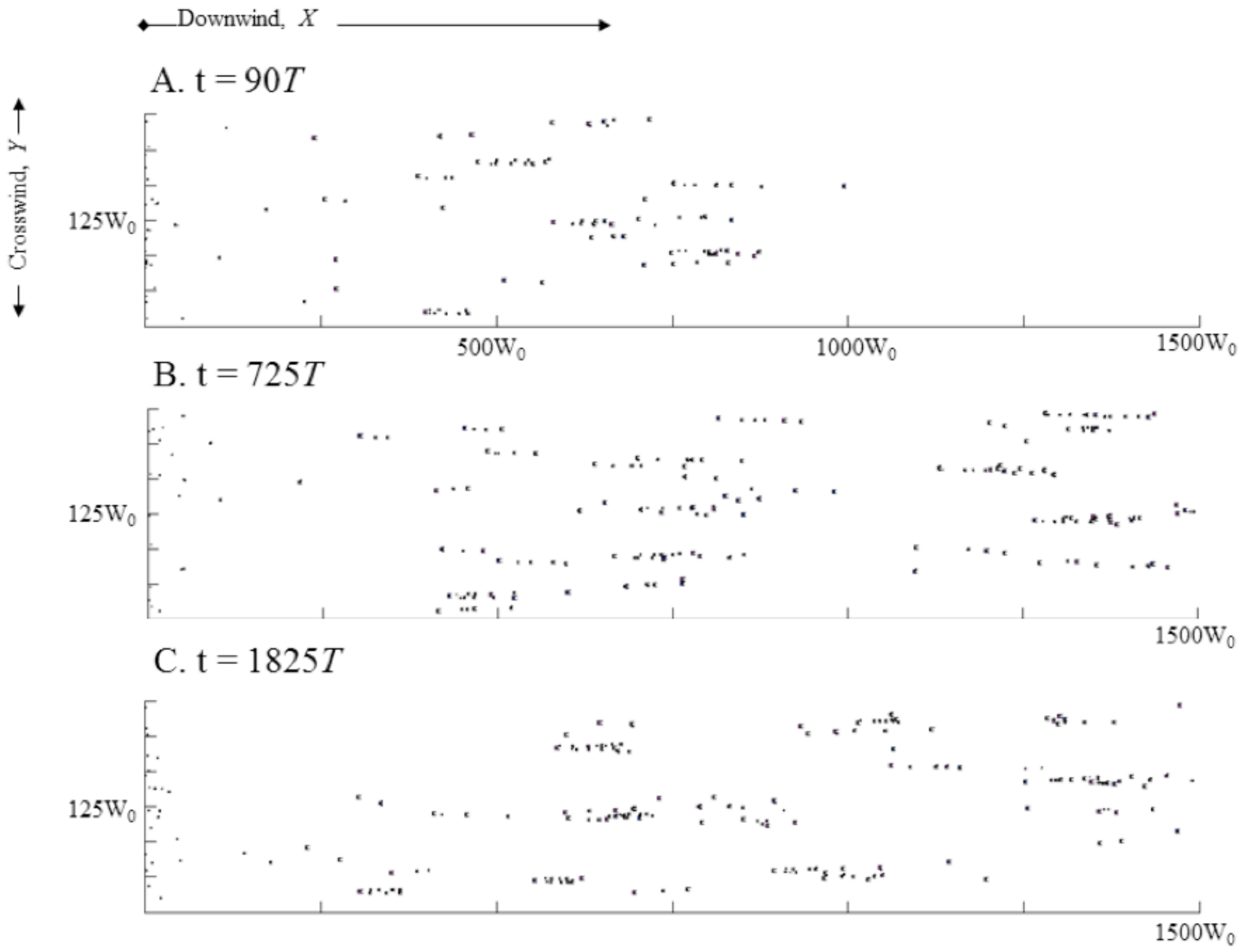}}
\caption{Time series revealing model approach and maintenance of a steady pattern for a lower-flux run, where $q_{b,i} = 0.20 q_{\rm sat}$.}
\label{S3}
\end{figure}
\begin{figure}[p]
\centerline{\includegraphics[width=\textwidth]{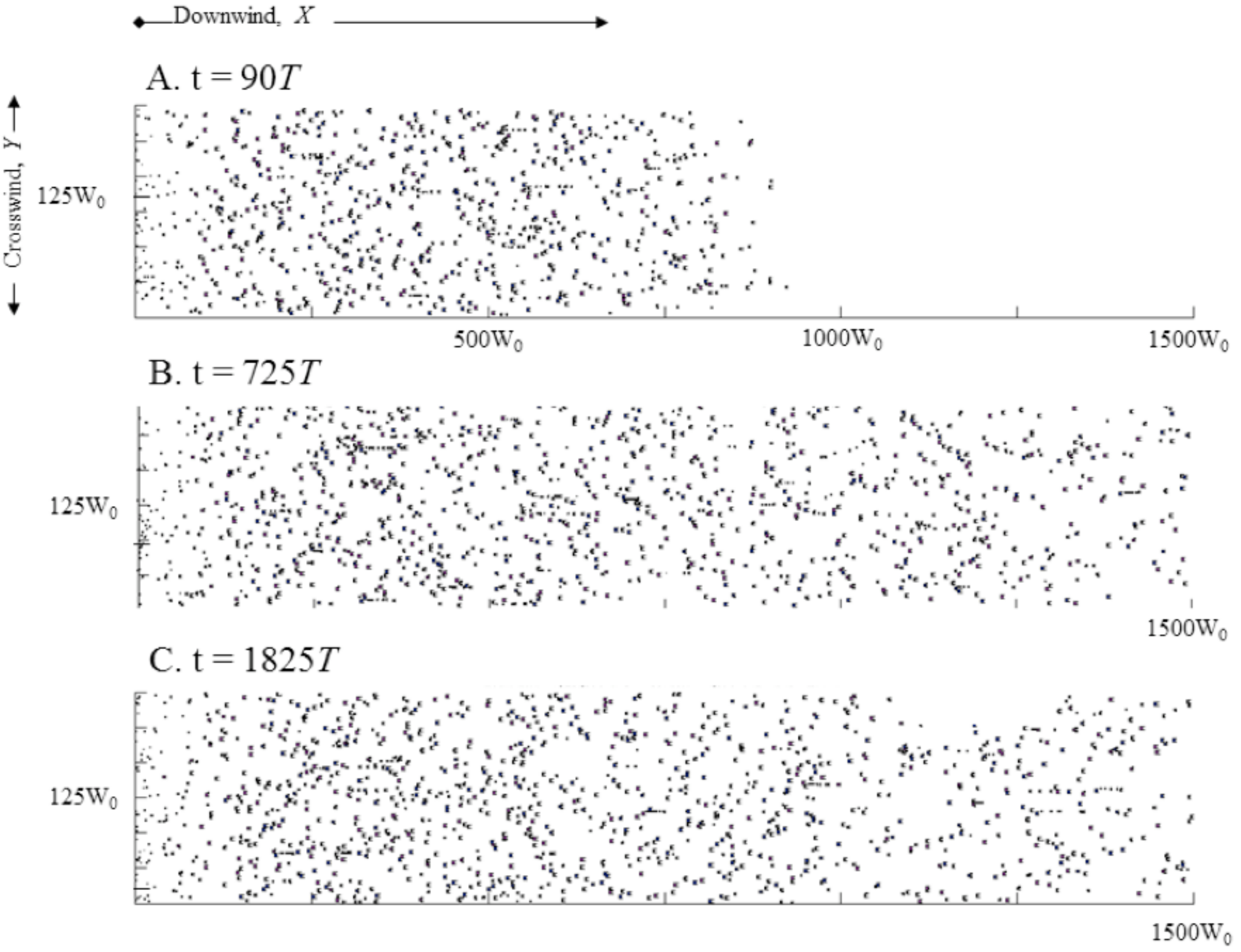}}
\centerline{\includegraphics[width=\textwidth]{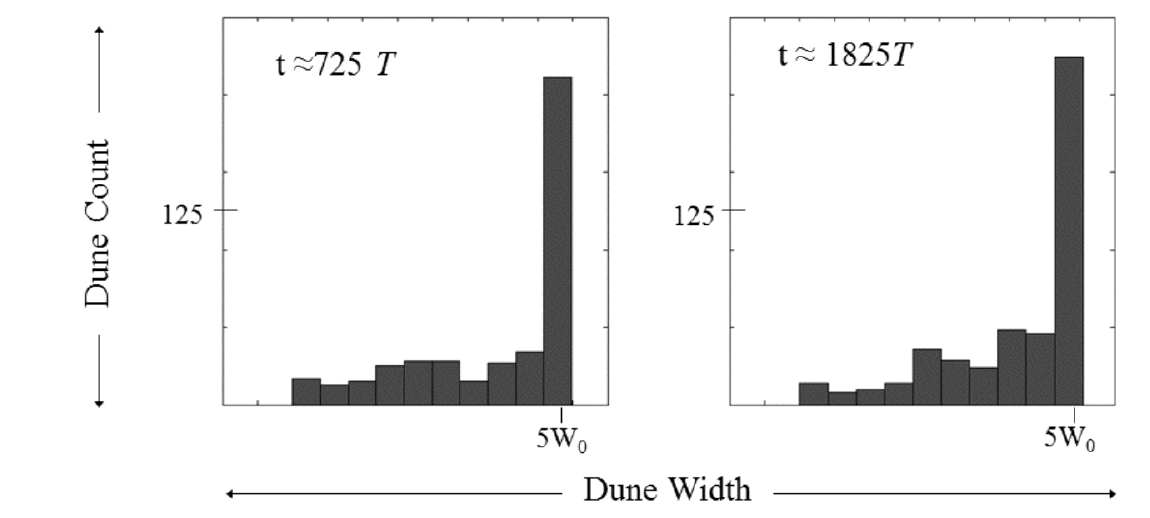}}
\caption{Time series and histograms of dune sizes (note one `characteristic' size) revealing model approach and maintenance of a steady pattern for a higher-flux run, where $q_{b,i} = 0.25 q_{\rm sat}$.}
\label{S4}
\end{figure}
\begin{figure}[p]
\centerline{\includegraphics[width=\textwidth]{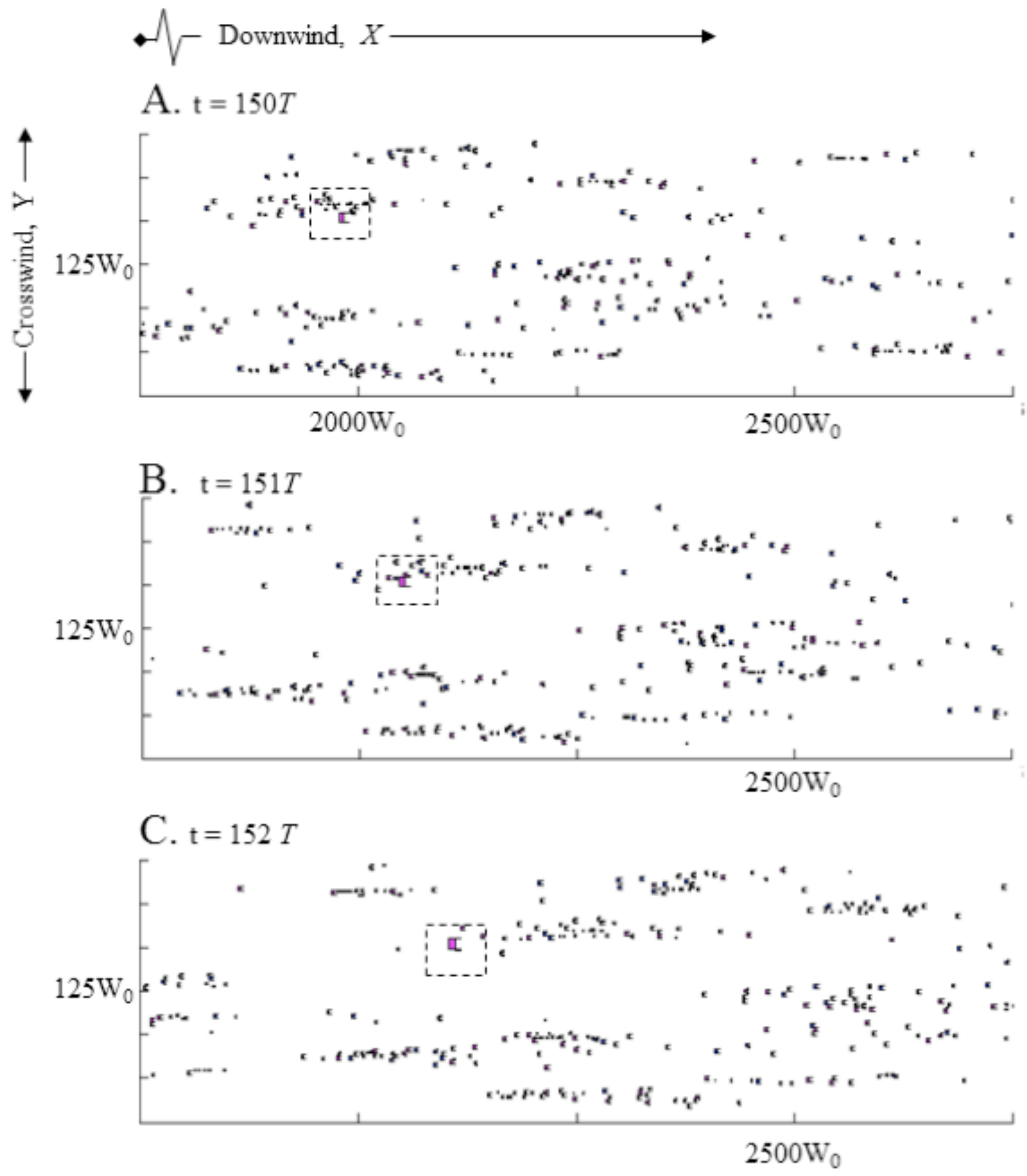}}
\caption{Example of stochastic and runaway growth of a mega-dune in a lower-flux run, where $q_{b,i} = 0.20 q_{\rm sat}$. Note this infrequent event happens far from the upwind boundary. The pink color indicates that the dune has gained sand since the previous time step.}
\label{S5}
\end{figure}
\begin{figure}[p]
\centerline{\includegraphics[width=\textwidth]{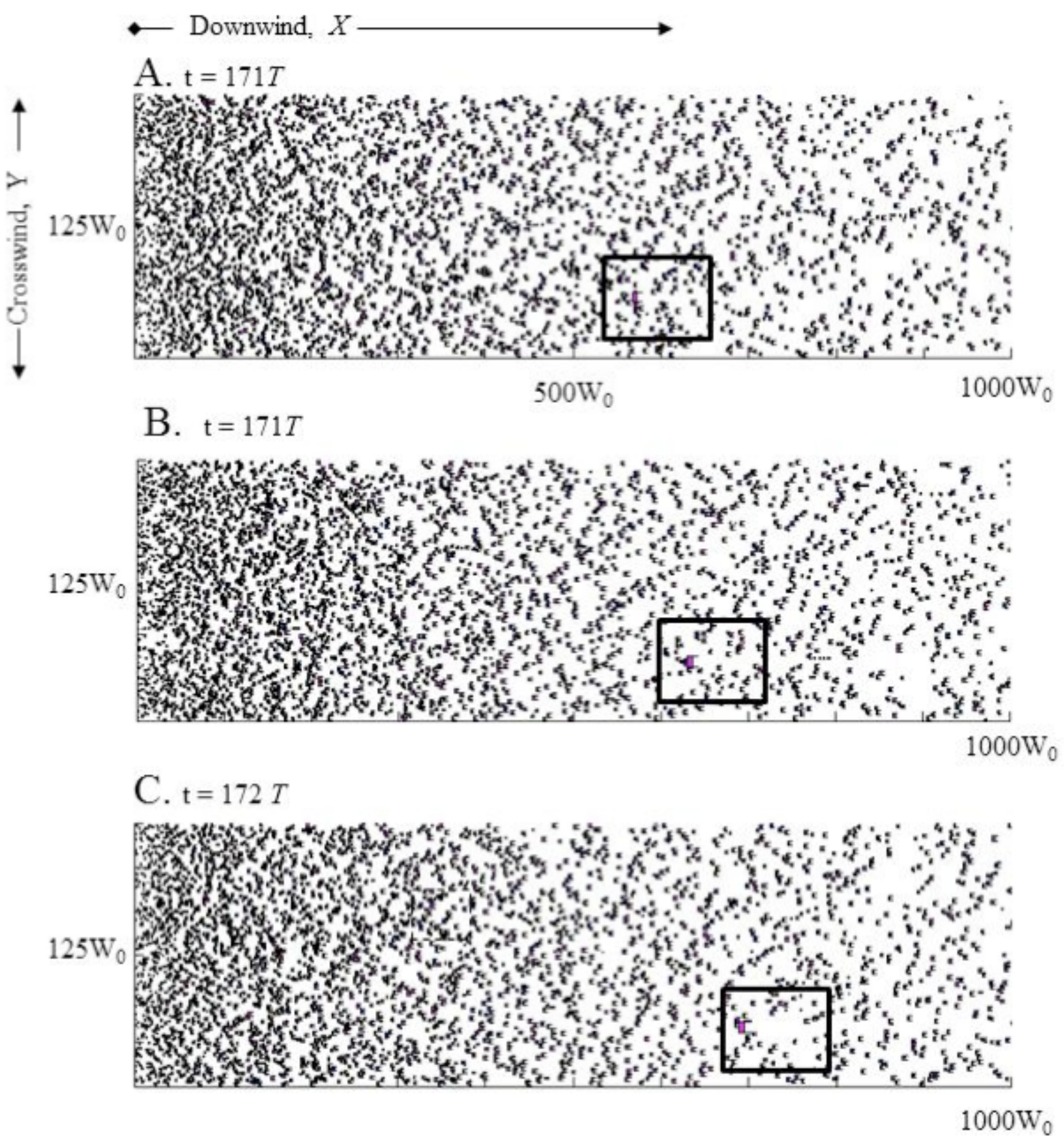}}
\caption{Example of stochastic and runaway growth of a mega-dune in a higher-flux run, where $q_{b,i} = 0.40 q_{\rm sat}$. The pink color indicates that the dune has gained sand since the previous time step.}
\label{S6}
\end{figure}
\begin{figure}[p]
\centerline{\includegraphics[width=\textwidth]{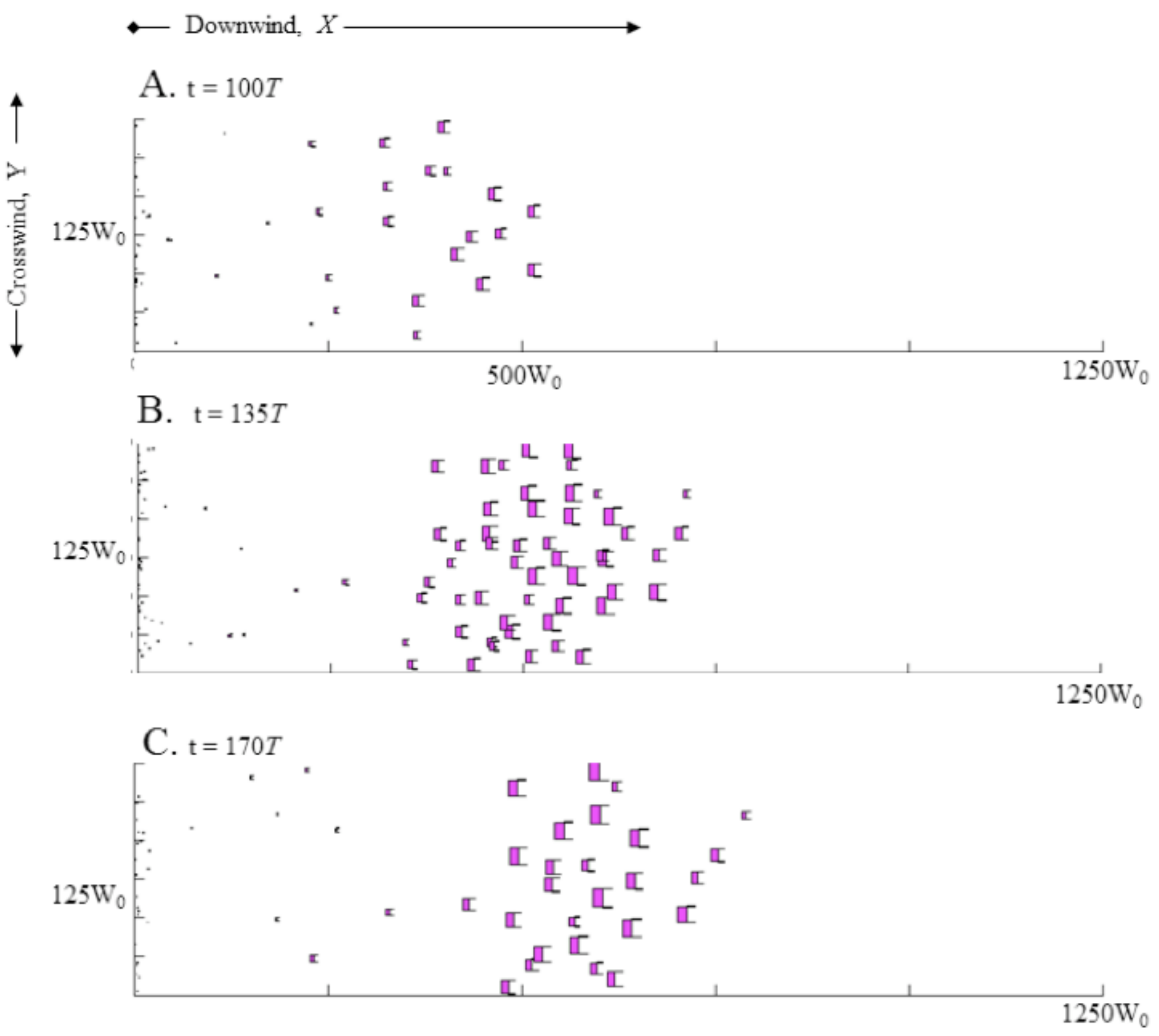}}
\caption{Disabling calving process leading to run away growth of all dunes on short time scales, example from lower-flux run, $q_{b,i} = 0.20 q_{\rm sat}$.}
\label{S7}
\end{figure}
\begin{figure}[p]
\centerline{\includegraphics[width=\textwidth]{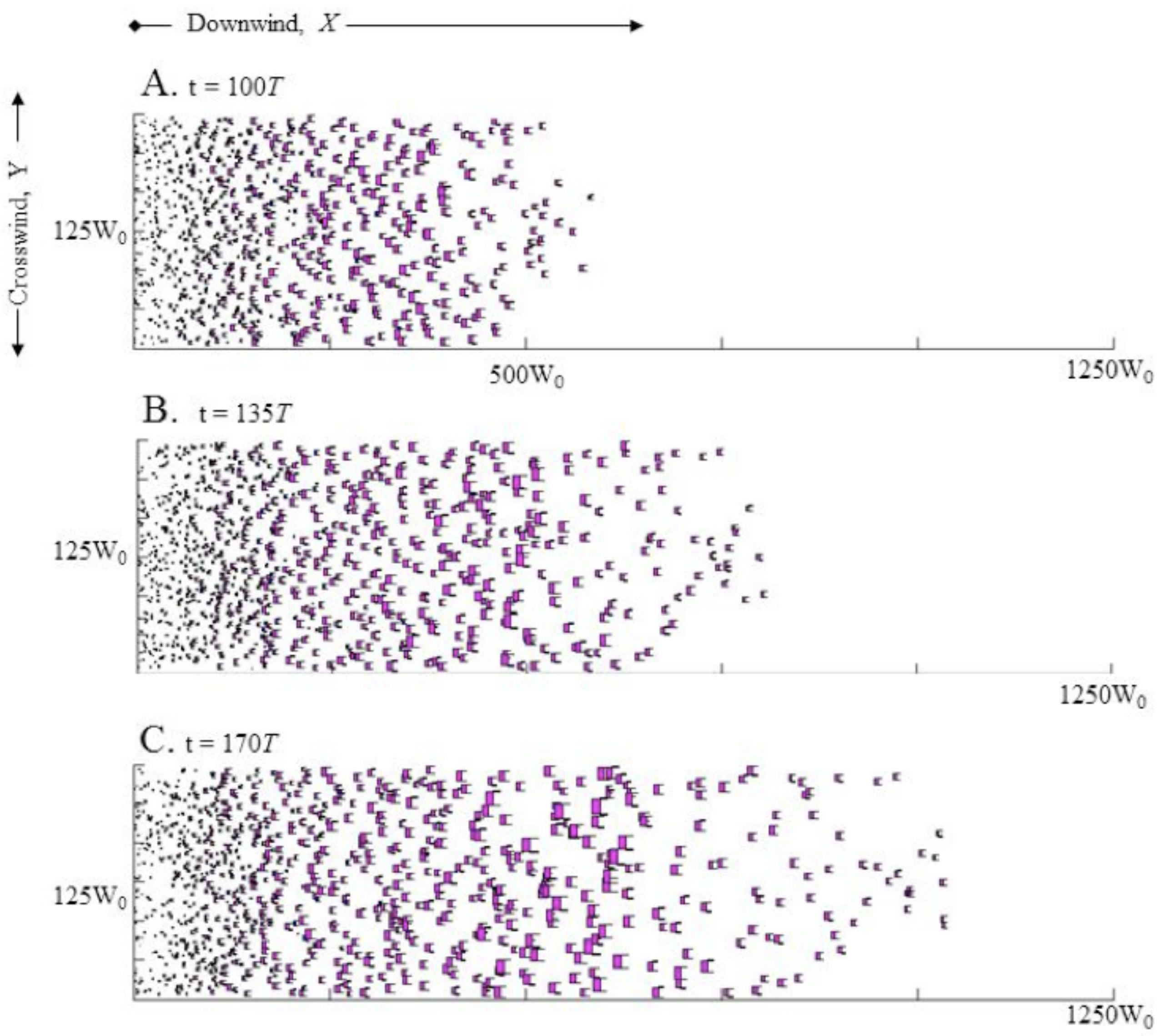}}
\caption{Disabling calving process leading to run away growth of all dunes on short time scales, example from higher-flux run, $q_{b,i} = 0.30 q_{\rm sat}$.}
\label{S8}
\end{figure}
\begin{figure}[p]
\centerline{\includegraphics[width=\textwidth]{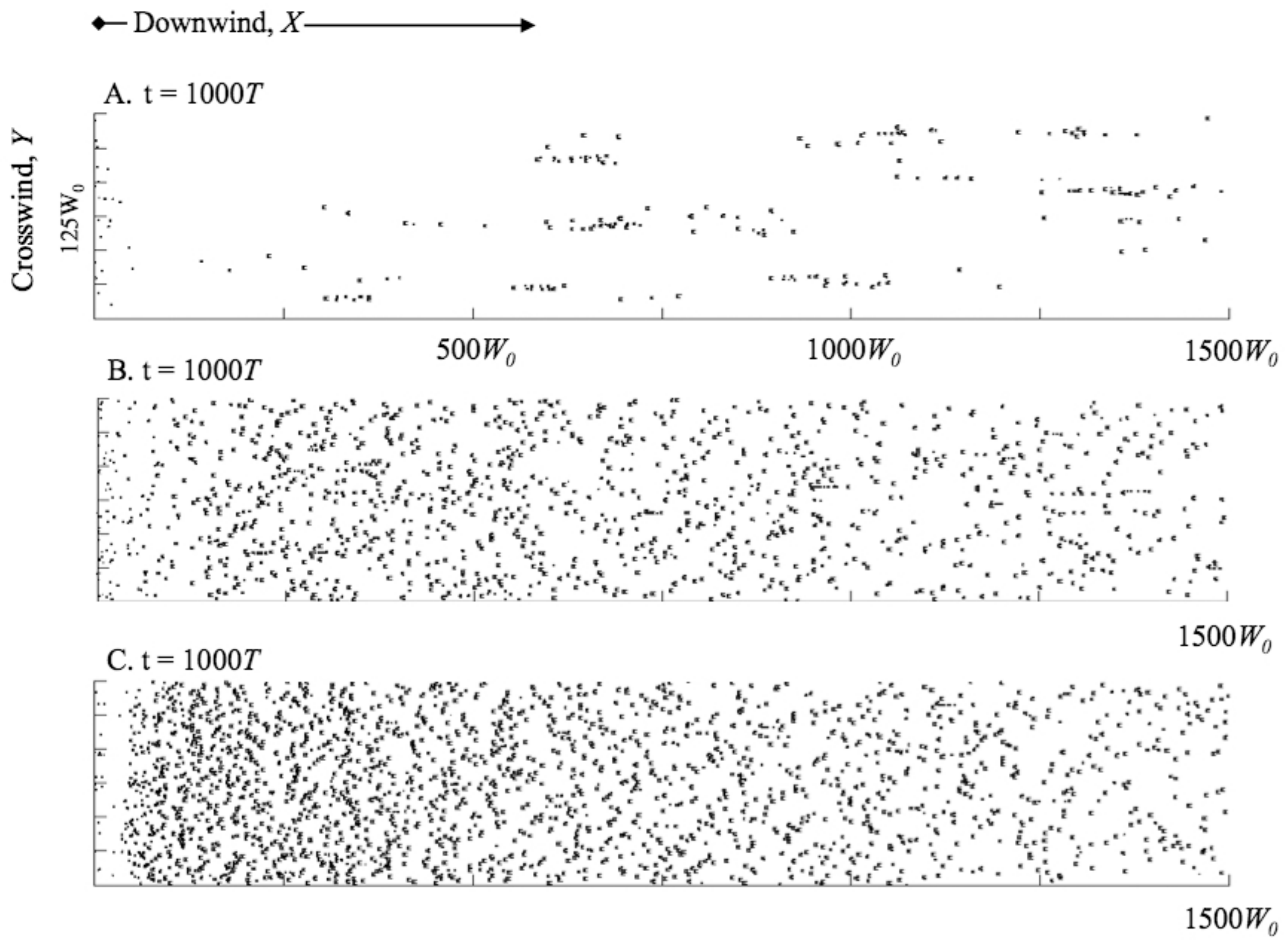}}
\caption{Visual appearance of steady state fields varies with the sand supply rate at the upwind boundary (A) $q_{b,i} = 0.20 q_{\rm sat}$, (B) $q_{b,i} = 0.25 q_{\rm sat}$ and (B) $q_{b,i} = 0.30 q_{\rm sat}$. As parameterized for the Tarfalya corridor, these results represent a $5$~km by $30$~km field after $\simeq 5000$~years.}
\label{S9}
\end{figure}
\begin{figure}[p]
\centerline{\includegraphics[width=\textwidth]{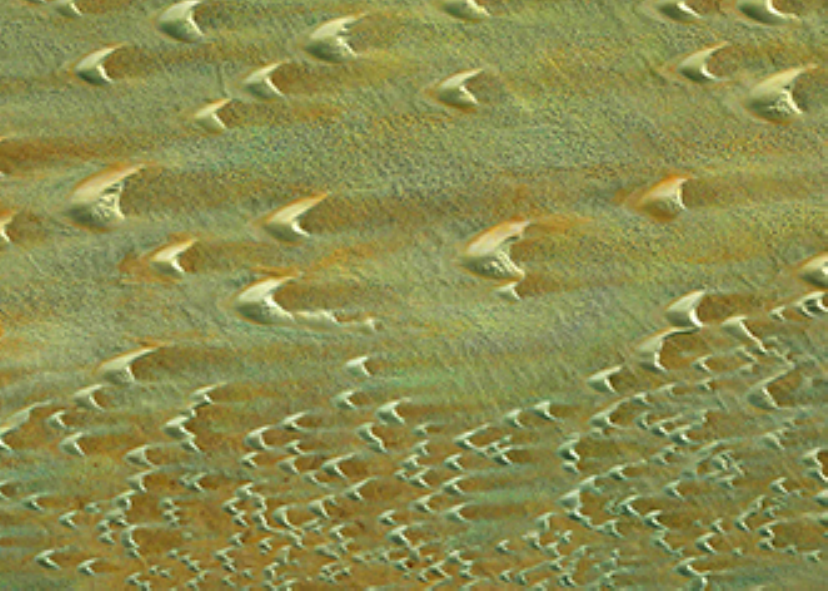}}
\caption{Structure in a natural field showing inverse relationship between dune size and density.}
\label{S10}
\end{figure}


\begin{thebibliography}{}

\bibitem{ACD02a} 
Andreotti B, Claudin P, Douady S (2002)
Selection of dune shapes and velocities (Parts 1 \& 2).
Eur Phys J B \textbf{28}, 321--352.

\bibitem{B07}
Baas ACW (2007)
Complex systems in Aeolian geomorphology.
Geomorphology \textbf{91}, 311--331.

\bibitem{DGB10}
Diniega S, Glasner K, Byrne S (2010)
Long-time evolution models of aeolian sand dune fields: Influence of dune formation and collision.
Geomorphology \textbf{121}, 55--68.

\bibitem{DPH10}
Dur\'an O, Parteli EJR, Herrmann HJ (2010)
A continuous model for sand dunes: Review, new developments and application to barchan dunes and barchan dune fields.
Earth Surf Process Landforms \textbf{35}, 1591--1600.

\bibitem{DSLH11}
Dur\'an O, Schwammle V, Lind PG, Herrmann HJ (2011)
Size distribution and structure of barchan dune fields.
Nonlinear Proc. Geophys. \textbf{18}, 455--467.

\bibitem{ENBK11}
Eastwood E, Nield J, Baas A, Kocurek G (2011)
Modeling controls on Aeolian dune-field pattern evolution.
Sedimentology \textbf{58}, 1391--1406.

\bibitem{E12}
Elbelrhiti H (2012)
Initiation and early development of barchans dunes:  A case study of the Moroccan Atlantic Sahara desert.
Geomorphology \textbf{138}, 181--188.

\bibitem{ECA05}
Elbelrhiti H, Claudin P , Andreotti B (2005)
Field evidence for surface wave induced instability of sand dunes.
Nature \textbf{437}, 720--723.

\bibitem{EAC08}
Elbelrhiti H, Andreotti B, Claudin P (2008)
Barchan dune corridors:  Field characterization and investigation of control parameters.
J Geophys Res \textbf{113}, F02S15.

\bibitem{EK10}
Ewing RC, Kocurek G (2010)
Aeolian dune-field pattern boundary conditions.
Geomorphology \textbf{114}, 175--187.  

\bibitem{GHCG12}
G\'enois M, Hersen P, Courrech du Pont S, Gr\'egoire G (2012)
When dunes move together, structure of deserts emerges.
\texttt{ArXiv:1211.7238}.

\bibitem{HAEACD04}
Hersen P, Andersen KH, Elbelrhiti H, Andreotti B, Claudin P, Douday S (2004)
Corridors of barchans dunes:  Stability and size selection.
Phys Rev E \textbf{69}, 011304.

\bibitem{HD05}
Hersen P, Douady S (2005)
Collision of barchans dunes as a mechanism of size regulation.
Geophys Res Lett \textbf{32}, L21403.

\bibitem{HB12}
Hugenholtz CH, Barchyn TE (2012)
Real barchan dune collisions and ejections.
Geophys. Res. Lett., \textbf{39}, L02306.

\bibitem{KNET05}
Katsuki A, Nishimori H, Endo HN, Taniguchi K (2005)
Collision dynamics of two barchan dunes simulated using a simple model.
J Phys Soc Japan \textbf{74}, 538--541.

\bibitem{KEM10}
Kocurek G, Ewing RC, Mohrig D (2010)
How do bedform patterns arise?  New views on the role of bedform interactions within a set of boundary conditions.
Earth Surf Process Landforms \textbf{35}, 51--63. 

\bibitem{KSH02}
Kroy K, Sauermann G, Herrmann HJ (2002)
Minimal model for aeolian sand dunes.
Phys Rev E \textbf{66}, 031302.

\bibitem{LSHK02}
Lima AR, Sauermann G, Herrmann HJ, Kroy K (2002)
Modeling a dune field.
Physica A \textbf{314}, 487--500.

\bibitem{ZNR10}
Zhang D, Narteau C, Rozier O (2010)
Morphodynamics of barchan and transverse dunes using a cellular automaton model.
J. Geophys. Res. \textbf{115}, F03041.

\end{thebibliography}
\end{document}